# Local communities obstruct global consensus:
# Naming game on multi-local-world networks

Yang Lou[1], Guanrong Chen[1], Zhengping Fan[2*], and Luna Xiang[3]

[1] Department of Electronic Engineering, City University of Hong Kong, Hong Kong SAR, China

[2] School of Data and Computer Science, Sun Yat-sen University, Guangzhou 510275, China

[3] Department of Applied Social Sciences, City University of Hong Kong, Hong Kong SAR, China

*Corresponding author: fanzhp@mail.sysu.edu.cn

**Abstract** – Community structure is essential for social communications, where individuals belonging to the same community are much more actively interacting and communicating with each other than those in different communities within the human society. Naming game, on the other hand, is a social communication model that simulates the process of learning a name of an object within a community of humans, where the individuals can generally reach global consensus asymptotically through iterative pair-wise conversations. The underlying network indicates the relationships among the individuals. In this paper, three typical topologies, namely random-graph, small-world and scale-free networks, are employed, which are embedded with the multi-local-world community structure, to study the naming game. Simulations show that 1) the convergence process to global consensus is getting slower as the community structure becomes more prominent, and eventually might fail; 2) if the inter-community connections are sufficiently dense, neither the number nor the size of the communities affects the convergence process; and 3) for different topologies with the same average node-degree, local clustering of individuals obstruct or prohibit global consensus to take place. The results reveal the role of local communities in a global naming game in social network studies.

## 1   Introduction

Individuals (or agents) employed in a naming game (NG) [1,2] are connected by a certain communication network. The network represents the relationships among involving agents, on which two agents can communicate directly with each other only if they are directly connected on the network. Isolated agent is not allowed in the underlying network, which is not participating the game and hence can be removed, thus information can be propagated to every agent so that the whole population may eventually reach global consensus (*i.e.*, convergence), in the sense that every agent keeps one and only one identical name to describe the object to be named. The convergence of NG may be observed via numerical simulations [1-3], proved theoretically [4], or verified empirically by humans-participated experiments [5]. As to the underlying communication network, the random-graph [6], small-world [7] and scale-free [8] networks are the most widely used ones for naming games [9-16], which will also be employed in the present study.

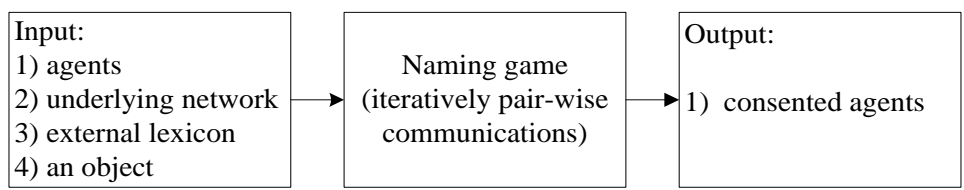

**Figure 1**   The framework of a minimal naming game.

Figure 1 shows the flowchart of a minimal NG, where *minimal* means that the model is defined with only the fundamental ingredients of the real-world lexicon propagation phenomenology. More complicated models can be further developed, if desired, based on this minimal version. The input of minimal NG includes: 1) a population of agents with empty memory, but each agent has infinite capacity of memory; 2) a connected underlying network indicating the relationships among the agents; 3) an infinite (or large enough) external lexicon which specifies a large number of different names; 4) an object (entity, idea, convention, or event, etc.) to be named by the population. The output is a population of consented agents, where every agent has one and only one identical name for the object in his memory. The convergence process will be recorded for analysis, in terms of *e.g.* the *number of total names* and the *number of different names* in the population, as well as the *success rate*. The changes with any input item will cause different converging features; for example, the case when all agents have a limited memory size [3].

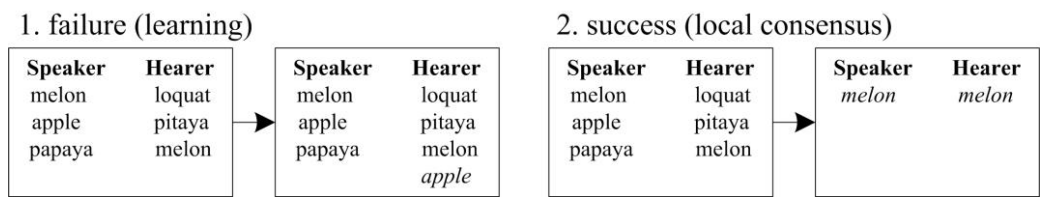

**Figure 2**    An example of one time step during pair-wise communication (two situations in total). The first situation is a failure or learning phase, where the hearer does not know the name *apple* before the speaker uttered it, so the hearer learns and keeps *apple* in his memory. The second is a success or local consensus, where the hearer has the speaker-uttered name *melon* in memory, so, as a result they both clear out all names except

keeping the name *melon*.

At each time step of the minimal NG, a pair of connected agents is randomly selected from the population, to be speaker and hearer respectively. If the object is unknown to the speaker, meaning that the speaker has no name in his memory to describe the object, then he will randomly pick a name from the external lexicon (which is equivalent to randomly invent a new name within the certain number of words in the lexicon), and then utters the name to the hearer. When the object is already known to the speaker, namely the speaker has one or several names in his memory, he will randomly pick a name from the memory and then utter it. After the hearer receives the name, he will search over his memory to see if he has the same name stored therein: if not, then he will store it into the memory; but if yes, then the hearer and the speaker reach consensus, so they both clear up all the names while keeping this common name in their respective memory. An example illustrating one time step of the pair-wise communication is given in Figure 2. This pair-wise success is referred to as local consensus hereafter. Such a pair-wise transmitting and receiving (or teaching and learning) process will continue to iterate until eventually the entire population of agents reach consensus, referred to as global consensus, meaning that all the agents agree to describe the object by the same name.

Each node of the underlying network represents an agent in NG, while each edge means that the two connected nodes can communicated to each other directly, in either pair-wise [9-16] or group-wise [17-19] communication setting. The number of connections of a node is referred to as its degree. The heterogeneity of social networks can generally be reflected by the scale-free networks [8,10,11], where a few agents have much larger degrees than most agents that have very small degrees. On the other hand, human communications are community-based, in the sense that people belonging to the same community are much more actively interacting and communicating with each other than those in different communities. Recall that the multi-local-world (MLW) model [20,21] is a kind of scale-free network, capable of capturing the essential features of many real-world networks with community structures. The degree distribution of the MLW network is neither in a completely exponential form nor in a completely power-law form, but is somewhere between them.

In particular, the MLW model shows good performance on capturing basic features of the Internet at the autonomous system (AS) level [22]. It is quite well known that human social networks also have AS-like structures. Therefore, it is quite reasonable to study a naming game of a population on an MLW communication network where a local world is a community formed not only by natural barriers such as mountains, rivers and oceans, but also by folkways, dialect and cultures. In this paper, therefore, naming game is studied under an MLW network framework, with three typical topologies of human communication networks, namely random-graph, small-world and scale-free networks, respectively.

Naming game is well-known to be simulation-based due to its large-scale and intrinsic-complexity nature. Most of the previous studies on NG show a feature of eventual convergence, even with small learning errors in communications [23]. It is observed that a population of $N$ nodes requires $O(N^{1.5})$ iterations to reach global consensus on fully-connected networks [12]. As for random-graph, small-world and scale-free networks, the order is $O(N^{1.4})$ [10,12,17]. However, it is observed in [2,24] that, when networks have prominent community structures, global consensus may be obstructed or even fail. But if a certain number of *committed agents* can be introduced in, global convergence can be regained [2], where a committed agent is one that has one and only one fixed name, who insists in his own name persistently. In [24], a differential equation method was employed to explain the non-convergence phenomenon in a bi-community network. In the present paper here, the effects on convergence is studied by varying the number and size of the local communities, and the focus is on the relationship of inter-connections and intra-connections as well as the clustering degree of the underlying networks. It can be observed from Figure 4 that the convergence time on MLW exceeds $N^{1.9}$ ($N = 1000$, $N^{1.9} = 5.01 \times 10^5$), meaning that when naming game is performed on an MLW network, the situation is quite different from those on other networks studied previously [10,12,17].

The main contributions of this study include the following findings: 1) the convergence process to global consensus is becoming slower as the community structure within a network becomes more prominent, and eventually might fail, where a prominent community structure means that the ratio of inter-community connections and intra-community connections is small; 2) if the inter-community connections are sufficiently dense, neither the number nor the size of the communities affects the convergence process; and 3) for different topologies with the same average node-degree, local clustering of individuals obstruct or prohibit global consensus to take place. The simulation results reveal the role of local communities in a global naming game in social networks.

The rest of the paper is organized as follows. In Section 2, the multi-local-world model is introduced, followed by extensive simulation results with analysis in Section 3. Finally, Section 4 concludes the investigation.

## 2    The Multi-local-world Networks

Here and throughout, all random operations (*e.g.*, random generation, selection, addition or deletion) follow a uniform distribution.

The algorithm for generating an MLW network [21] with $N$ nodes can be summarized as follows.

The initialization starts with $N_{LW}$ isolated local-worlds. Within each local-world, there are $m_0$ nodes connected by $e_0$ edges. At each time step, a value $r$ $(r \in (0,1))$ is generated at random.

***a.***    If $0 < r < p_1$, perform addition of a new local-world of $m_0$ nodes connected by $e_0$ edges, which is added to the existing network.

***b.***    If $p_1 \leq r < p_2$, perform addition of a new node to a randomly selected local-world $LW$ by preferential attachment: the new node is added to the selected local-world, establishing in $e_1$ new connections (edges). The new node is connected to $e_1$ nodes existing in the local-world according to the following preferential probability:

$$\Pi(k_i) = \frac{k_i + \alpha}{\sum_{j \in LW}(k_j + \alpha)} \qquad (1)$$

where $k_i$ is the degree of node $i$ within the local-world $LW$ and $\alpha$ is a tunable parameter.

***c.***    If $p_2 \leq r < p_3$, perform addition of edges within a randomly selected local-world $LW$: $e_2$ edges are added to this $LW$. For each new edge, one end is connected to a randomly picked node within the $LW$, while the other end is connected to a node selected also from the same $LW$ according to a probability $\Pi(k_i)$ given by Eq. (1). This process repeats $e_2$ times.

***d.***    If $p_3 \leq r < p_4$, perform deletion of edges within a randomly selected local-world $LW$: $e_3$ edges are deleted from $LW$. The purpose is to remove more edges that connect to small-degree nodes. To do so, randomly select a node from $LW$. Remove the edges of this node one by one, according to the following probability where $k_i$ is the degree of the node at the other end of the edge:

$$\Pi'(k_i) = \frac{1}{N_{LW} - 1} \cdot (1 - \Pi(k_i)) \qquad (2)$$

where $N_{LW}$ is the number of nodes within the $LW$ and $\Pi(k_i)$ is given by Eq. (1). This process repeats $e_3$ times.

***e.***    If $p_4 \leq r < 1$, perform addition of edges among local-worlds: $e_4$ edges are added to connect different local-worlds. First, two different local-worlds are picked at random. Then, one node is selected within each local-world according to the probability given by Eq. (1). An edge is finally added between these two nodes. This process repeats $e_4$ times.

The initial number of nodes is $N_{LW} \cdot m_0$ and the termination number is $N > N_{LW} \cdot m_0$ (typically, much larger). The generation algorithm stops when totally $N$ nodes have been generated into the network.

Note that throughout the above process, the generation of repeated connections, self-loops and isolated nodes should be avoided or removed. The detailed generating algorithm of MLW networks as well as the calculation of its degree distribution can be found in [21]. As shown above, there are totally eleven tunable parameters, among which only two parameters are of interest in the present paper, *i.e.*, the number of local-worlds $N_{LW}$ and the initial number $m_0$ of nodes within each local-world.

According to [24], it is hard for a population to reach globally consensus if the underlying network has multiple communities. The underlying network used in [24] is a combination of several scale-free networks, where the combination is generated by a reversed preferential attachment probability. Specifically, the intra-connections within each community are based on a preferential attachment probability given by Eq. (1), while the inter-connections between communities are generated according the following preferential attachment probability:

$$\Pi(k_i) = \frac{1/k_i + \alpha}{\sum_{j \in LW}(1/k_j + \alpha)} \tag{3}$$

Only bi-community and tri-community networks are studied in [24]. In a bi-/tri-community network, all the inter-community links are actually connecting to the other one/two communities. In this case, one community may either converge locally or be distracted by another community, which could be considered as a single source of interference, or two. However, on an MLW network, there are many sources of interference affecting the local convergence of each community, and as a result the situation is much more intrinsic and complicated.

In this paper, the MLW model introduced above will be employed, in which both the number $N_{LW}$ and the initial size $m_0$ are tunable parameters. By simply adjusting these two parameters, the NG can be performed on a set of more generalized networks with multiple communities, more realistic to represent the real human society and language development therein.

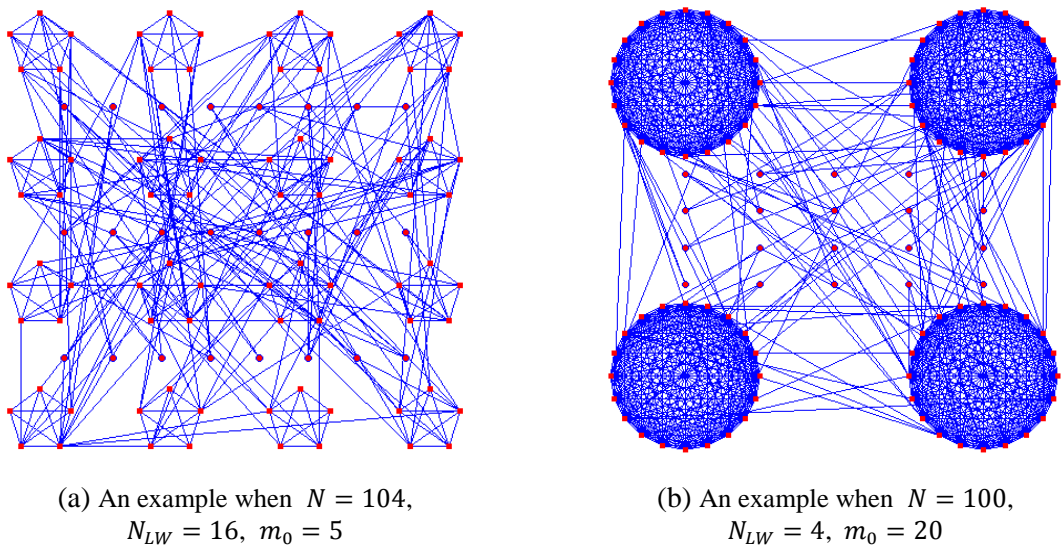

(a) An example when $N = 104$, $N_{LW} = 16$, $m_0 = 5$

(b) An example when $N = 100$, $N_{LW} = 4$, $m_0 = 20$

**Figure 3**   Two examples of multi-local-world network. The red squares represent the initial nodes assigned to the initial local-worlds, while the red dots with blue rims are the nodes being added afterwards. Since $e_0 = m_0 \cdot (m_0 - 1)/2$, all the local-worlds are fully-connected initially, and some edges may be removed by operation $d$ with the probability of 0.04 as indicated in Table 1.

## 3    Results and Analysis

The minimal NG is studied on MLW networks for it simulates the Internet as well as many social networks realistically. There are mainly eleven parameters, among which we are interest in only two, *i.e.*, the number of local-worlds $N_{LW}$ and the initial number $m_0$ of nodes within each local-world. The other nine out of eleven parameters are fixed, as set in [21], which are $p_1$, $p_2$, $p_3$, $p_4$, $e_0$, $e_1$, $e_2$, $e_3$, and $e_4$. Their values are presented in Table 1, along with their correspondence or meanings of such parameter settings. All the initial local-worlds are fully-connected, so the parameter $e_0 = m_0 \cdot (m_0 - 1)/2$, some of links will be removed by operation $d$ yet some will be added back by operation $c$. Other than $N_{LW}$ and $m_0$, a change on the rest nine parameters leads to nothing but different forms of the underlying MLW network, *e.g.*, changing $p_3$ will alter the probability of adding back links within some local-worlds.

Next, denote the number of individuals (population size) by $N$, which satisfies $N > N_{LW} \cdot m_0$ otherwise there will be only $N_{LW}$ isolated local-worlds, so the network is not connected [21]. Introduce a new parameter $\rho$ $(0 < \rho < 1)$, as the rate of initially assigned nodes in the local-worlds: when $\rho = 0$, there is no local-world and the network degenerates to a scale-free one since every node is added by a preferential attachment; when $\rho = 1$, it generates several isolated local-worlds without any additional nodes or edges. The purpose of introducing $\rho$ is to change the above inequality to be equality:

$$\rho \cdot N = N_{LW} \cdot m_0 \qquad (4)$$

**Table 1** Parameter values and their correspondence or meanings

| Parameter Setting | Meaning |
|---|---|
| $p_1 = 0$ | Operation *a* (addition of new local-worlds) is not performed |
| $p_2 = 0.28$ | Operation *b* (addition of a new node to a local-world) is performed with probability 0.28 |
| $p_3 = 0.39$ | Operation *c* (addition of edges within a local-world) is performed with probability 0.11 (= 0.39 − 0.28) |
| $p_4 = 0.43$ | Operation *d* (deletion of edges within a local-world) is performed with probability 0.04 (= 0.43 − 0.39); meanwhile, operation *e* (addition of edges among local-worlds) is performed with probability 0.57 (= 1.00 − 0.43) |
| $e_0 = m_0 \cdot (m_0 - 1)/2$ | Initially, local-worlds are isolated but in each of them the nodes are fully-connected |
| $e_1 = e_2 = e_3 = e_4 = 2$ | At each time step, when operations *b*, *c*, and *d* are performed, the number of edges added or deleted is 2 |

The comparing simulation is carried out by varying $\rho$, $m_0$, and $N_{LW}$. *Convergence time* will be used as the measure, which refers to the number of time steps when global convergence is reached. In the following comparisons, 1) $\rho$ is fixed and the *convergence time* affected by the dynamics of the number and size of local-worlds is examined; 2) the *convergence time* is studied when the rate $\rho$ is varying, with fixed values of $m_0$ and $N_{LW}$; and 3) the convergence progresses of MLW networks built on three typical models are compared; *i.e.*, random-graph (RG) [6], small-world (SW) [7] and scale-free (SF) [8] networks.

The population size is set and fixed to $N = 1000$, and the cases when the population size is 500 and 1500 are studied in the supplementary information (SI) [25]. The maximum number of iteration is set to $10^7$ and data are collected from 30 independent runs and then averaged. Here, $10^7$ iterations are empirically large enough for this study. Also, denote the number of different names at time step $t$ by $N_{diff}(t)$. Simulation shows that

$$1 \le N_{diff}(10^7) \le N_{LW} \qquad (5)$$

When $N_{diff}(10^7) = 1$, it has reached global convergence, while when $1 < N_{diff}(10^7) \le N_{LW}$, it means the local-worlds have converged to different names, respectively, as can be seen in Table 2. In addition, with a long time period $\tau \gg 0$, one has that

$$N_{diff}(10^7 - \tau) = N_{diff}(10^7) \qquad (6)$$

which means that the number of different words is not changed during a long time. Note also that $N_{diff}$ is monotonically non-increasing in this converging (or converged) stage. The parameter $\tau$ represents the length of the stagnation that one can observe from, *e.g.*, the curves in Figures 7 and 8. Empirically, by observing a long time period of $\tau \gg 0$, the convergence process makes no more progress; thus, one may consider it as sufficiently well converged. Considering the conditions shown in Eqs. (5) and (6) together, by setting the maximum number of iteration to $10^7$, one can see that the population has converged sufficiently well.

**Table 2** The number of total words at iteration $10^7$ comparing to the number of local-worlds. As $m_0$ is set to 26 different values, the number of local-worlds is calculated by $N_{LW} = \lfloor \rho N / m_0 \rfloor$. It can be seen that the parameter setting that always yields convergence (in 30 independent runs) corresponds to $N_{diff}(10^7) = 1$; otherwise, $N_{diff}(10^7) > 1$. Putting together all the cases, one has $1 \leq N_{diff}(10^7) \leq N_{LW}$; especially, when $m_0 \geq 30$, $N_{diff}(10^7)$ is approaching $N_{LW}$, meaning that every local-world converges to one different name.

| | $m_0$ | 3 | 4 | 5 | 6 | 7 | 8 | 9 | 10 | 11 | 12 | 13 | 14 | 15 |
|---|---|---|---|---|---|---|---|---|---|---|---|---|---|---|
| $\rho$ | $N_{diff}(10^7)$ | 1 | 1 | 1 | 1 | 1 | 1 | 1 | 1 | 1 | 1 | 1 | 1 | 1 |
| = 0.5 | $N_{LW}$ | 166 | 125 | 100 | 83 | 71 | 62 | 55 | 50 | 45 | 41 | 38 | 35 | 33 |
| $\rho$ | $N_{diff}(10^7)$ | 1.1 | 1 | 1 | 1 | 1 | 1 | 1 | 1 | 1 | 1 | 1 | 1 | 1.4 |
| = 0.7 | $N_{LW}$ | 233 | 175 | 140 | 116 | 100 | 87 | 77 | 70 | 63 | 58 | 53 | 50 | 46 |
| | | | | | | | | | | | | | | |
| | $m_0$ | 16 | 17 | 18 | 19 | 20 | 30 | 40 | 50 | 60 | 70 | 80 | 90 | 100 |
| $\rho$ | $N_{diff}(10^7)$ | 1 | 1 | 1 | 3.2 | 5.7 | 15.0 | 11.7 | 9.7 | 7.9 | 6.9 | 5.8 | 4.9 | 4.9 |
| = 0.5 | $N_{LW}$ | 31 | 29 | 27 | 26 | 25 | 16 | 12 | 10 | 8 | 7 | 6 | 5 | 5 |
| $\rho$ | $N_{diff}(10^7)$ | 1.5 | 7.9 | 11.7 | 22.4 | 25.4 | 22.8 | 17.0 | 13.9 | 11.0 | 10.0 | 8.0 | 6.9 | 7.0 |
| = 0.7 | $N_{LW}$ | 43 | 41 | 38 | 36 | 35 | 23 | 17 | 14 | 11 | 10 | 8 | 7 | 7 |

### 3.1 Convergence time vs the number and size of local-worlds

The number $m_0$ of initial nodes of each local-world are set to 26 different values: varying from 3 to 19 with an increment 1, and from 20 to 100 with an increment 10, to have different scenarios. The rate of initially assigned nodes is set to $\rho = 0.5$ and 0.7, respectively, as shown in Figures 4(a) and (b). It can be seen from Figure 4 that relatively small sizes of communities, despite a large number of communities, do not hinder global convergence. Since nodes are sufficiently connected inside and outside various communities, prominent intra-community connections, on the one hand facilitate local consensus, but on the other hand make the global convergence more difficult, especially when some other communities had already converged or almost converged to different words respectively. For reference, the ratio of inter-links versus intra-links for each node is plotted in Figure 4. Since the population size is fixed to be 1000, as $m_0$ increases the ratio becomes smaller.

In the box plot shown in Figure 4, the blue box indicates that the central 50% data lie in this section; the red bar is the median value of all 30 datasets; the upper and lower black bars are the greatest and least values, excluding outliers which are represented by the red pluses.

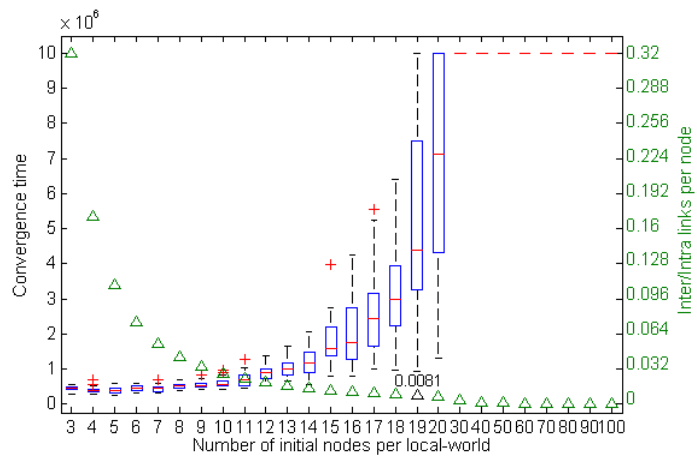

(a)   $\rho = 0.5$

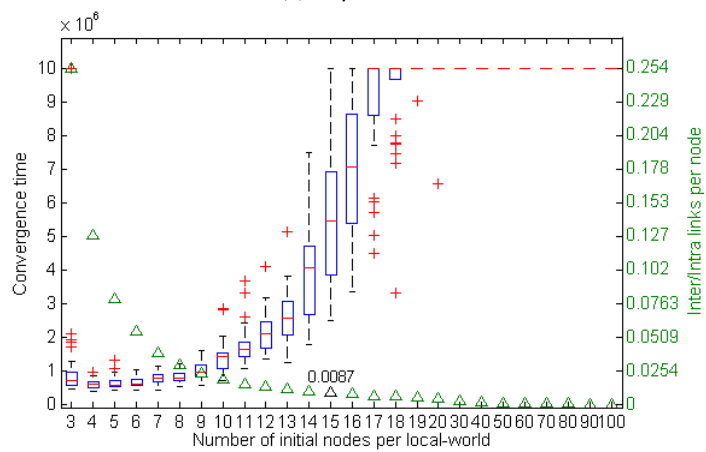

(b)   $\rho = 0.7$

**Figure 4** The box plot of the convergence time vs the initial nodes in each local-world, $m_0$, with (a) $\rho = 0.5$ and (b) $\rho = 0.7$. The number of local-worlds can be calculated by Eq. (4), and since it should be an integer, $N_{LW} = \lfloor \rho N/m_0 \rfloor$, where $\lfloor x \rfloor$ is the largest integer less than or equal to $x$. The mean value of convergence time in both figures is concave: it first slightly decreases when $m_0$ increases from 3 to 5, and then increases as $m_0$ continues to increase. When $m_0 = 4$ and 5, it converges in the fastest speeds in both cases. In (a), when $m_0 = 19$, it starts to show non-converged behaviors, and the according mean ratio of inter-connection versus intra-connection is 0.0081. As for (b), when $m_0 = 15$, it starts to show non-converged behaviors, and the according mean ratio is 0.0087.

Table 3 shows the mean ratio of inter-connection versus intra-connection per node, when the population starts to show occasionally non-converged behaviors. The cases of $0.4 \leq \rho \leq 0.7$ is studied here, because when $\rho$ is very small ($\rho \leq 0.3$), the initial community structure is unclear, and actually it generates a preferential attached network rather than an MWL. In contrast, when $\rho$ is greater than $0.8$, the generated network may be disconnected due to the lack of inter-connections. The $\bar{R}$ values are calculated by the averaged ratio of the number of inter-connections divided by the number of intra-connections within each community, and then divided by the number of nodes within the same community. For example, in Figure 5(b), the nodes $n1$, $n2$, $n3$ and $n4$ cluster to become a community, which has 6 intra-connections and 1 inter-connection, thus the ratio for each node of this community is $(1/6)/4 = 0.0417$. The $\bar{R}$ value is the mean value of the ratios over all the communities within the underlying network.

**Table 3** The mean ratio ($\bar{R}$) of inter-connection versus intra-connection per node, when the population starts to show occasionally non-converged behaviors. Here, $m_0^{min}$ is the minimum value of $m_0$ when the population starts to become non-converged. When $m_0 = m_0^{min} - 1$, it is always converged, *e.g.*, the cases of $\rho = 0.5$ and $\rho = 0.7$ can also be observed from Figure 4 where, when $m_0 = 19$ and $15$ respectively, it starts to show non-converged behaviors. Moreover, $N_{LW} = \lfloor \rho N/m_0 \rfloor$ and $\bar{R}$ is the mean ratio of inter-connection versus intra-connection for each node, averaged from 30 independent runs, and $Std$ is the standard deviation.

|  | $m_0^{min}$ | $N_{LW}$ | $\bar{R}$ | $Std$ |
|---|---|---|---|---|
| $\rho = 0.4$ | 20 | 20 | 0.0084 | $4.63 \times 10^{-4}$ |
| $\rho = 0.5$ | 19 | 26 | 0.0081 | $3.83 \times 10^{-4}$ |
| $\rho = 0.6$ | 17 | 35 | 0.0084 | $4.09 \times 10^{-4}$ |
| $\rho = 0.7$ | 15 | 46 | 0.0087 | $6.07 \times 10^{-4}$ |

Now, examine Table 3 more closely. First, as the $\rho$ value increases, the $m_0^{min}$ value decreases, meaning that when more nodes are initially allocated in the communities, these nodes should be more sparsely distributed in different small communities, rather than gathering in just a few large communities; otherwise, the global convergence may be hindered. Second, the ratio ($\bar{R}$) of inter-link versus intra-links stays relatively stable in different cases, meaning that on average such values give an approximate lower bound to sufficiently many inter-community connections towards global convergence.

Table 4 shows the average degrees, average path lengths and average clustering coefficients of all the generated MLW networks. It shows that as $m_0$ increases, both the average degree and average clustering coefficient increase, while the average path length decreases. This means that, on average, when $m_0$ increases, the networks are better connected, yet more clustered. Better connections (greater average degree and shorter average path length) facilitate convergence in NG [17,18], while local clustering and forming communities hinder convergence. At the extreme, one can assume that any sub-network in a fully-connected network is a local community. In this case, both intra-community and inter-community connections are maximized, thus there is no barrier existing amongst the communities in a fully-connected network. Barriers preventing communities from global convergence are formed only if the intra-community connections are strong while the inter-community connections are weak.

**Table 4**    The feature statistics of all the multi-local-world networks in simulation. Here, $\langle k \rangle$ is the average degree, $\langle pl \rangle$ is the average path length and $\langle cc \rangle$ is the average clustering coefficient. As $m_0$ increases, both $\langle k \rangle$ and $\langle cc \rangle$ increase, while $\langle pl \rangle$ decreases.

| $m_0$ | | 3 | 4 | 5 | 6 | 7 | 8 | 9 | 10 | 11 | 12 | 13 | 14 | 15 |
|---|---|---|---|---|---|---|---|---|---|---|---|---|---|---|
| | $\langle k \rangle$ | 6.09 | 6.82 | 7.24 | 7.58 | 8.09 | 8.56 | 9.20 | 9.54 | 9.95 | 10.42 | 11.05 | 11.80 | 11.97 |
| $\rho = 0.5$ | $\langle pl \rangle$ | 3.98 | 3.81 | 3.74 | 3.70 | 3.66 | 3.61 | 3.56 | 3.53 | 3.52 | 3.49 | 3.46 | 3.37 | 3.42 |
| | $\langle cc \rangle$ | 0.33 | 0.35 | 0.36 | 0.40 | 0.42 | 0.43 | 0.44 | 0.49 | 0.52 | 0.52 | 0.53 | 0.52 | 0.57 |
| | $\langle k \rangle$ | 4.37 | 5.42 | 5.98 | 6.72 | 7.26 | 8.01 | 8.57 | 9.41 | 10.01 | 10.61 | 11.66 | 12.09 | 12.78 |
| $\rho = 0.7$ | $\langle pl \rangle$ | 5.21 | 4.56 | 4.48 | 4.24 | 4.23 | 4.08 | 4.02 | 3.90 | 3.87 | 3.84 | 3.64 | 3.70 | 3.63 |
| | $\langle cc \rangle$ | 0.41 | 0.41 | 0.47 | 0.51 | 0.55 | 0.57 | 0.61 | 0.62 | 0.65 | 0.68 | 0.66 | 0.70 | 0.72 |
| $m_0$ | | 16 | 17 | 18 | 19 | 20 | 30 | 40 | 50 | 60 | 70 | 80 | 90 | 100 |
| | $\langle k \rangle$ | 12.55 | 13.06 | 13.04 | 13.91 | 14.56 | 19.00 | 23.78 | 29.78 | 33.64 | 39.00 | 43.34 | 45.94 | 54.61 |
| $\rho = 0.5$ | $\langle pl \rangle$ | 3.38 | 3.35 | 3.40 | 3.33 | 3.31 | 3.17 | 3.07 | 2.96 | 2.92 | 2.87 | 2.84 | 2.84 | 2.77 |
| | $\langle cc \rangle$ | 0.58 | 0.60 | 0.61 | 0.64 | 0.65 | 0.73 | 0.78 | 0.79 | 0.82 | 0.86 | 0.86 | 0.86 | 0.90 |
| | $\langle k \rangle$ | 13.48 | 14.19 | 14.83 | 15.61 | 16.43 | 22.91 | 29.80 | 37.18 | 42.32 | 51.39 | 54.36 | 59.83 | 72.41 |
| $\rho = 0.7$ | $\langle pl \rangle$ | 3.58 | 3.56 | 3.49 | 3.42 | 3.42 | 3.29 | 3.14 | 3.09 | 2.97 | 2.91 | 2.79 | 2.75 | 2.72 |
| | $\langle cc \rangle$ | 0.72 | 0.74 | 0.75 | 0.74 | 0.77 | 0.84 | 0.86 | 0.90 | 0.89 | 0.93 | 0.91 | 0.92 | 0.94 |

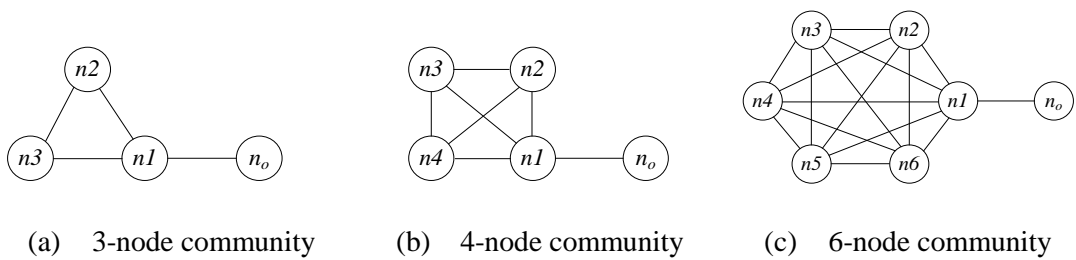

   (a)   3-node community      (b)   4-node community      (c)   6-node community

**Figure 5**    An example illustrating intra-community and inter-community connections. Here, $n_o$ is a node outside a community. The ratio of intra-connections vs inter-connections is: (a) 3:1; (b) 6:1; and (c) 15:1.

Figure 5 shows an example illustrating how the intra-connections become stronger when the community size increases. As can be seen from the figure, the ratio of intra-connections vs inter-connections is getting smaller as the community size is getting larger (see Figure 5(a) 3:1, Figure 5(b) 6:1, and Figure 5(c) 15:1). If one wishes to keep the ratio constant, *e.g.*, 3:1, then for a 4-node community there should be 2 nodes being connected externally, while for a 6-node community the number of inter-connections should be 5. Note that the number of inter-community connections is fixed. The inter-connections are generated during the addition of $N \cdot (1 - \rho)$ nodes, repeatedly by randomly selecting operations from *a* to *e* (defined in Section 2). As shown in Figure 5, if the number of inter-community connection is fixed, while the size of fully-connected community is growing, then the number of external connections is becoming insufficient for convergence.

In a nutshell, the inter-community connections of MLW networks should be kept constant, and the number and size of communities should be changed (reducing the number of communities and enlarging the size of each community), so that intra-community connections are getting stronger. As a result, as intra-connections increase, while inter-connections are kept constant, the convergence process will be slowed down and eventually failed. This explains more clearly the incremental convergence time shown in Figure 4.

### 3.2 Convergence time vs the rate of initially assigned nodes

In this section, both the number and size of local-worlds are fixed, while the rate $\rho$ of initially assigned nodes is varied from 0.1 to 0.9. As can be seen from Figure 6, a common pattern is that, when $\rho$ is small enough (*i.e.*, $\rho \leq 0.6$ in Figure 6(a); $\rho \leq 0.5$ in Figure 6(b); $\rho \leq 0.3$ in Figure 6(c)), different values of $\rho$ do not affect the convergence time at all. Note that the inter-community connections are generated during the addition of $N \cdot (1 - \rho)$ nodes. When $\rho$ is small enough for certain networks, this means that the inter-community connections are substantial and probably sufficient already to achieve global convergence.

As $\rho$ continues to increase, $N \cdot (1 - \rho)$ deceases, thus the inter-community connections are reducing and become insufficient if $\rho$ reaches certain large values (*i.e.*, when $\rho > 0.6$ in Figure 6(a); $\rho > 0.5$ in Figure 6(b); $\rho > 0.3$ in Figure 6(c)). Denote the threshold value by $\rho_{th}$. Then, when $\rho \leq \rho_{th}$, the convergence time is not affected by $\rho$, while when $\rho > \rho_{th}$, the convergence time increases drastically as $\rho$ increases. As can be seen from Figure 6, as $m_0$ increases ($m_0 = \{4, 10, 18\}$), $\rho_{th}$ decreases ($\rho_{th} = \{0.6, 0.5, 0.3\}$). This phenomenon can also be observed when the population size is 500 and 1500, respectively [25]. This phenomenon can be explained by the example shown in Figure 5, where the number of intra-community connections is $\binom{m_0}{2}$. This means that when $m_0$ is small, the number of intra-community connections is relatively small, so that the required inter-community connections become fewer, thus $\rho$ does not affect the convergence time, until that number becomes relatively large. In contrast, when $m_0$ is large, the number of intra-community connections is relatively large, thus even $\rho$ is relatively small, the convergence time is clearly affected, due to the large number requirement of inter-community connections.

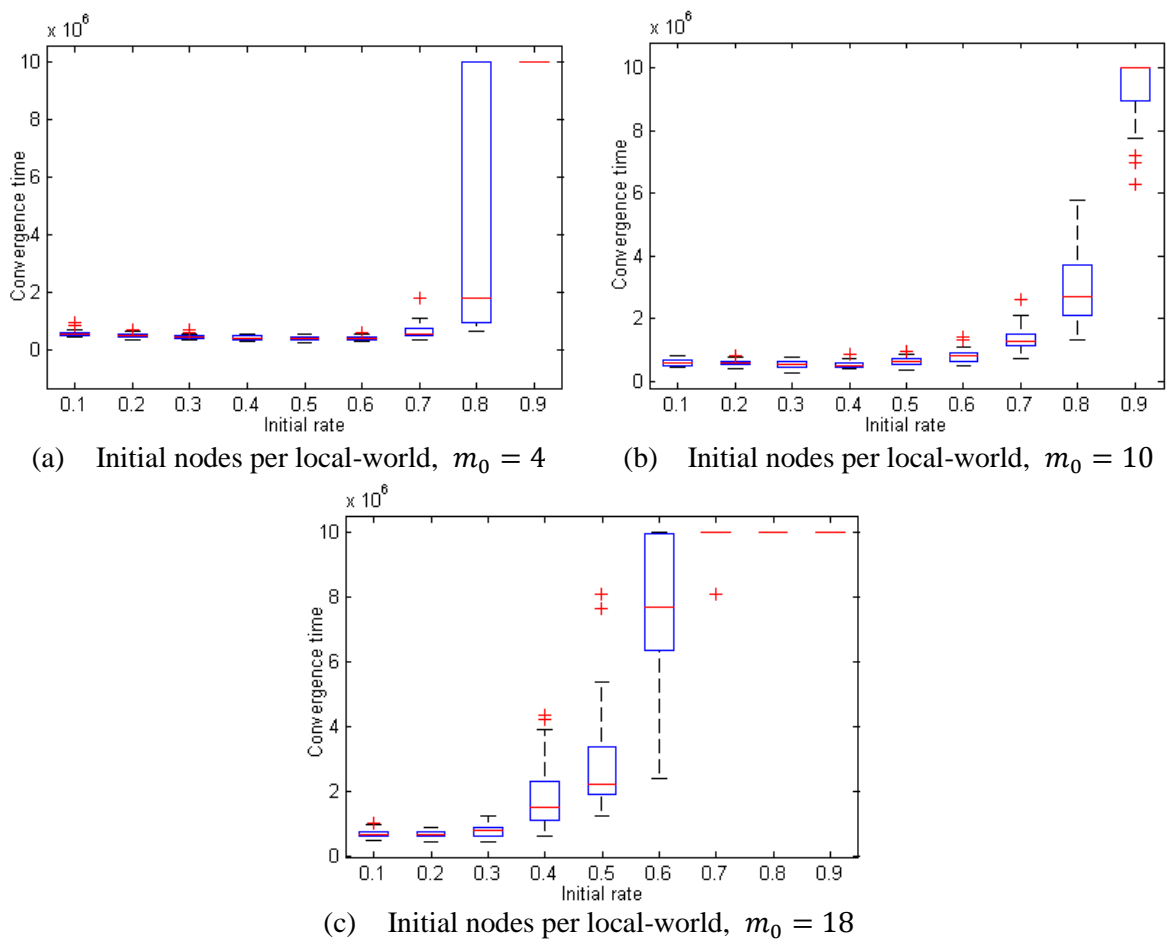

(a)   Initial nodes per local-world,  $m_0 = 4$          (b)   Initial nodes per local-world,  $m_0 = 10$

(c)   Initial nodes per local-world,  $m_0 = 18$

**Figure 6**    The box plot of the convergence time vs the rate  $\rho$  of initially assigned nodes. The number of initial nodes in each local-world  $m_0$  is set to 4, 10, and 18, respectively, the number of local-worlds is calculated by  $N_{LW} = \lfloor \rho N / m_0 \rfloor$ . In each figure, as  $\rho$  is varied from 0.1 to 0.9, the convergence time shows oscillations slightly prior to a prominent ascending progress.

### 3.3   Convergence process

The convergence progress of MLW networks are compared on three typical topology networks, *i.e.*, random-graph (RG) [6], small-world (SW) [7] and scale-free (SF) [8] networks. The comparison is implemented by the convergence progress in terms of the number of total words, the number of different words and the success rate. For fairness and also for convenience, four sets of data are chosen, with  $m_0$ = 10, 20, 30 and 100, on which the average degrees of the MLW networks are 9.41, 16.43, 22.91, 72.41, respectively. These data values are used as the connecting probabilities, exactly for generating RG networks and approximately for generating SW and SF networks. The feature statistics of the generated networks are summarized in Table 5, together with the statistics of the MLW for reference.

**Table 5** The feature statistics of the four networks. Here, $\langle k \rangle$ is the average degree, $\langle pl \rangle$ is the average path length and $\langle cc \rangle$ is the average clustering coefficient. The data for MLW is collected from experiments in Section 3.1, while those for the RG, SW and SF networks are generated using the $\langle k \rangle$ values of MLW for reference. As a result, the four types of networks have very similar $\langle k \rangle$ values.

| Reference | $\langle k \rangle$ | | | | $\langle pl \rangle$ | | | | $\langle cc \rangle$ | | | |
|---|---|---|---|---|---|---|---|---|---|---|---|---|
| | MLW | RG | SW | SF | MLW | RG | SW | SF | MLW | RG | SW | SF |
| $m_0 = 10$ | **9.41** | 9.51 | 10.00 | 9.96 | **3.90** | 3.32 | 3.86 | 2.95 | **0.62** | 0.01 | 0.35 | 0.05 |
| $m_0 = 20$ | **16.43** | 16.54 | 16.00 | 15.92 | **3.42** | 2.75 | 3.21 | 2.67 | **0.77** | 0.02 | 0.38 | 0.06 |
| $m_0 = 30$ | **22.91** | 22.95 | 22.00 | 21.85 | **3.29** | 2.55 | 2.84 | 2.50 | **0.84** | 0.02 | 0.38 | 0.07 |
| $m_0 = 100$ | **72.41** | 72.36 | 72.00 | 70.63 | **2.72** | 1.93 | 2.04 | 1.95 | **0.94** | 0.07 | 0.39 | 0.16 |

As shown in Table 5, four types of networks have very similar values of average degrees. However, MLW has the longest average path length and the highest clustering coefficient values. SW is with the second longest average path length and the second highest clustering coefficient values. Both RG and SF have smaller values on these two features.

In Figures 7, 8 and 9, the four cases of different parameter settings are: (a) $\langle k \rangle \approx 9.41$, (b) $\langle k \rangle \approx 16.43$, (c) $\langle k \rangle \approx 22.91$, and (d) $\langle k \rangle \approx 72.41$, and these four types of networks of the same (or similar) average degree are compared in the same figure for clarity.

In Figure 7, the four sub-figures in the figure share two common phenomena: 1) the population with underlying network RG converges the fastest, followed by SF, SW, and MLW. MLW only converges in the case in Figure 7(a), but does not converge in the cases shown in Figures 7(b), (c), and (d); 2) the curves with underlying network RG has the highest peak, followed by SF and SW, and MLW holds the lowest. As also shown in Table 5, RG has the smallest clustering coefficient values, followed by SF and SW, and MLW has the greatest, meaning that MLW has strongest tendency in clustering and forming communities than the other three networks. SW has relatively strong tendency in clustering. This leads to the following two observations: 1) in the early stage, individuals within communities reach convergence easily and quickly, as can be seen from Figure 9, where the larger the clustering coefficient is, the faster the success rate value increases, in the early stage; also in Figure 7, the number of total words decreases fast in the early stage, when there is a prominent community structure; 2) in the late stage, the inter-community convergence process is delayed or even prevented by the multi-community structure. This can be further summarized as follows: given the same average degree, the less clustered network has a convergence curve with a higher peak and sharper decline; while the more clustered network has a flatter curve with a lower peak.

Note that when the underlying network is a tree (with average degree $2 - \frac{2}{N}$ and clustering coefficient zero) or a globally-fully-connected network (with average degree $N - 1$ and clustering coefficient one), these two extreme cases are not investigated in the above simulations because, in these two special cases, for a given the average degree value the clustering coefficient cannot be adjusted.

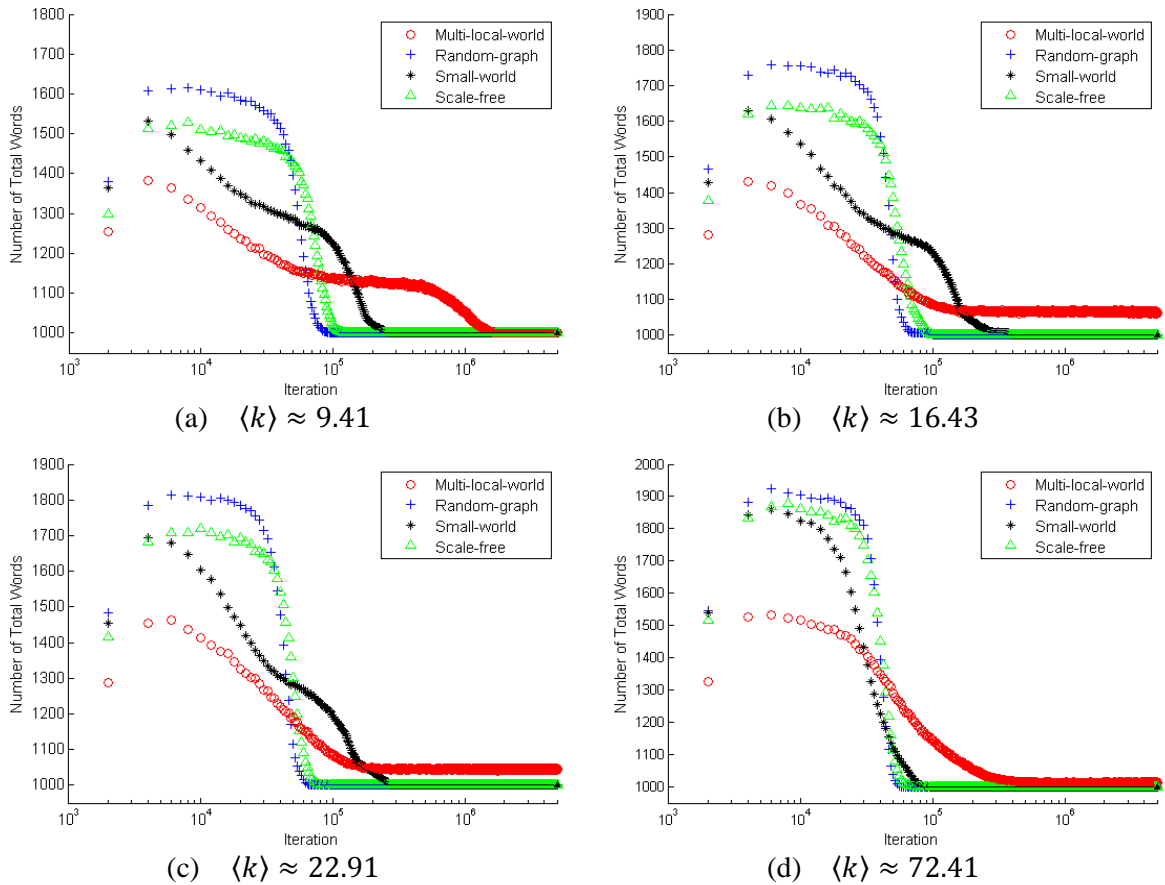

(a)  $\langle k \rangle \approx 9.41$

(b)  $\langle k \rangle \approx 16.43$

(c)  $\langle k \rangle \approx 22.91$

(d)  $\langle k \rangle \approx 72.41$

**Figure 7**   Comparison of the convergence processes in terms of the number of total words. In each figure, RG (blue plus) curve has the highest peak and the fastest speed to reach convergence, while MLW (red circle) has the lowest peak and it converges gradually. The curves of SW (black star) and SF (green triangle) behave between the curves of RG and MLW.

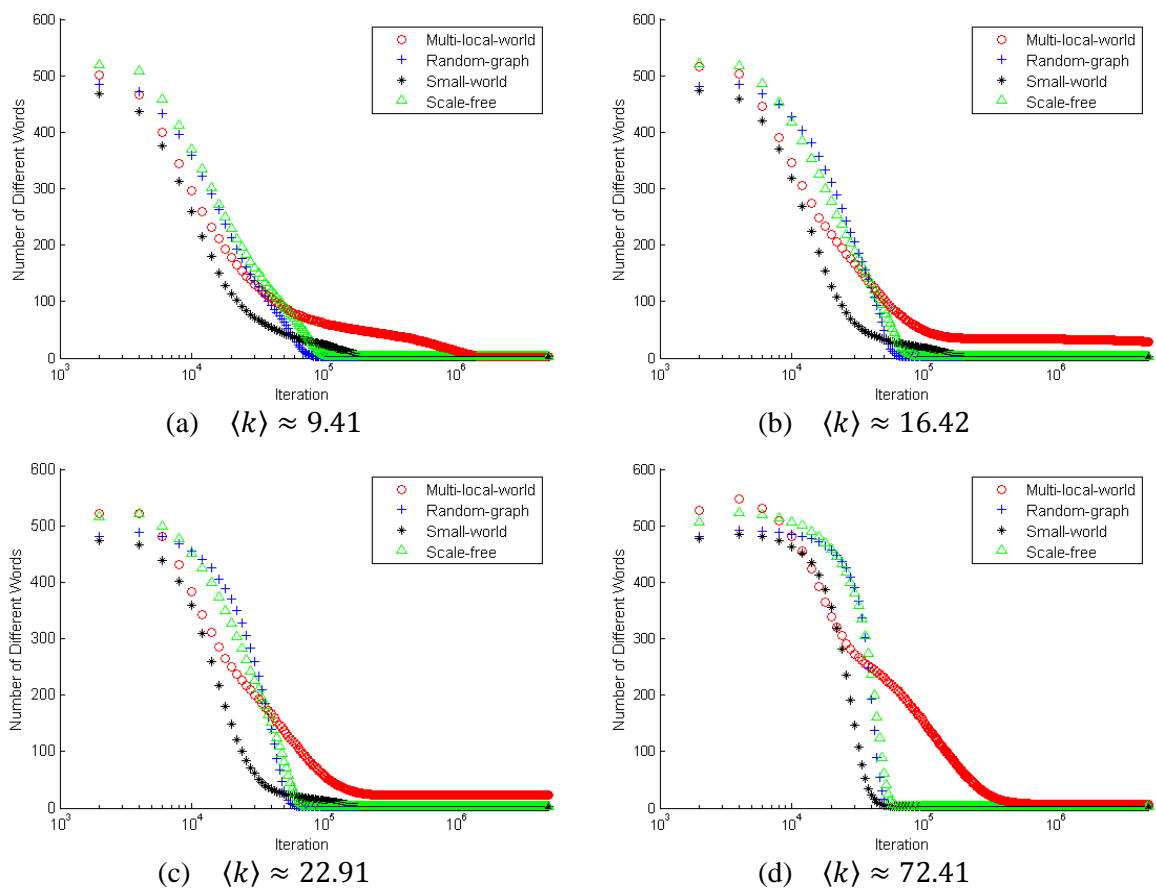

**Figure 8** Comparison of the convergence processes in terms of the number of different words. Differing from what are shown in Figure 7, the peaks of all four curves in each sub-figure are similar to each other. Similarly to Figure 7, the rank of convergence speeds is: RG (blue plus) converges the fastest; SW (black star) ranks the second, followed by SF (green triangle); MLW (red circle) converges the slowest.

In Figure 8, although the ranking of the convergence is exactly the same as what is shown in Figure 7, the peaks of the curves are similar to each other. This is because not only the lexicon but also the game rules are identical for all types of underlying networks, namely, if the picked speaker has nothing in his memory then he randomly picks a name from the external lexicon.

Figure 9 shows the success rate. It is obvious that when a network has a small clustering coefficient value, its success rate curve is generally smooth. However, for SW and MLW networks, high clustering coefficient values generate very rough success rate curves. For SW, although rough, the success rate can eventually reach 1.0; but for MLW, if the population does not converge as shown in Figures 7 and 8, the success rate cannot reach 1.0. This is because, in the late stage: 1) individuals in intra-communities have already reached convergence, so that the success rate of intra-communication is as high as one, and 2) individuals in inter-communities have converged to generally different names, so that the success rate of inter-communication is likely to be as low as zero. As a result, the curves are fluctuating and visually fuzzy.

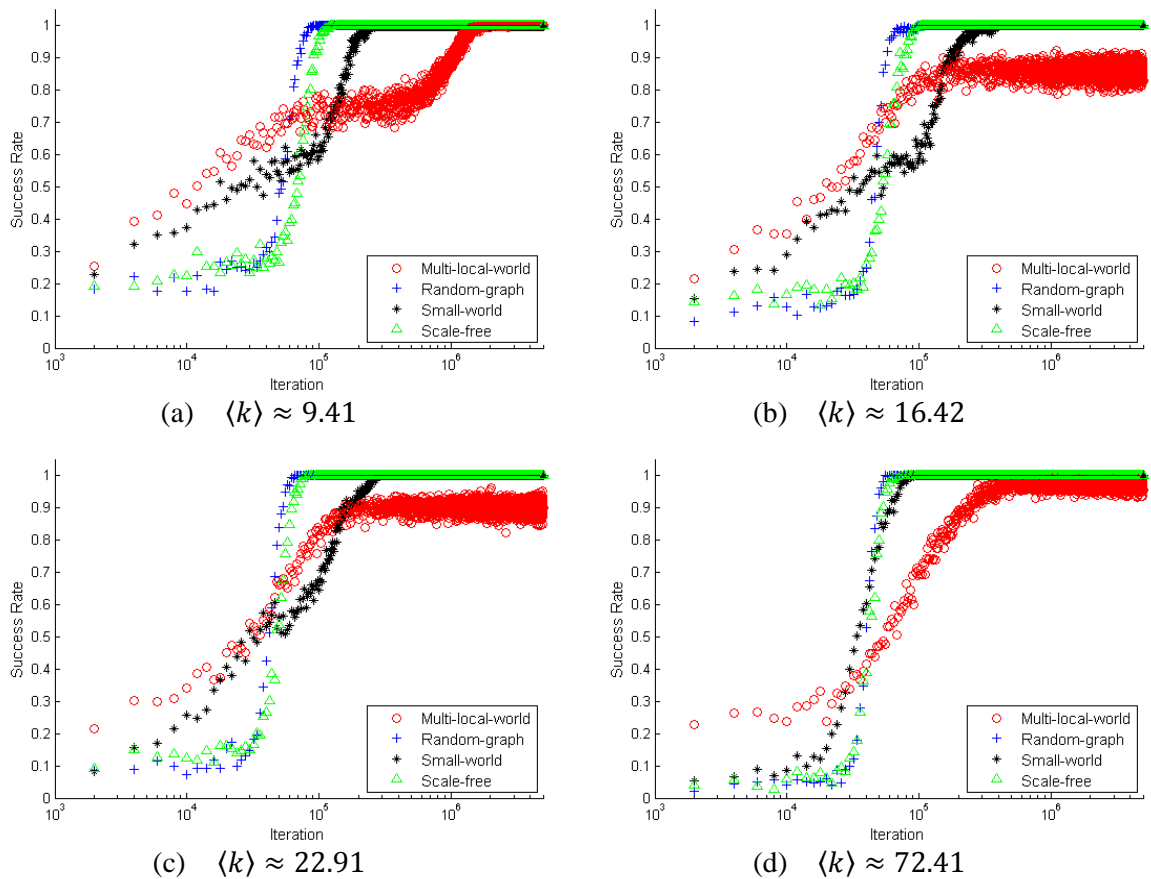

(a)   $\langle k \rangle \approx 9.41$

(b)   $\langle k \rangle \approx 16.42$

(c)   $\langle k \rangle \approx 22.91$

(d)   $\langle k \rangle \approx 72.41$

**Figure 9**   Comparison of the convergence processes in terms of the success rate. The curves of RG and SW are simple, while the other two are fluctuating. In particular, the curves of MLW are visually fuzzy and do not eventually reach value 1, in all sub-figures (b), (c) and (d).

### 3.4   Discussions

Consider a real-life situation that there are two types of local communities: one located in a suburb of a metropolis (denoted by $LW_m$), and the other is a primitive tribe (denoted by $LW_p$). The $LW_m$ has many connections to the metropolis (as well as the world outside $LW_m$) such as road paths, telephone systems and the Internet, while an $LW_p$ has probably only one trail to go outside without any other communicating connections. Within both communities, people know each other therefore they have direct communications.

Considering the above scenario, the first and second experimental studies show that if the size of a community is relatively small, no matter it is an $LW_m$ or $LW_p$, information can be easily delivered to each individual within the community, so that they are affected by the outside world (and finally reach global consensus). However, if the community size is big, a large number of external links are required. Otherwise, many individuals cannot receive information from outside, and hence the community (*e.g.*, an $LW_p$ of large size) can only reach local convergence, rather than global convergence.

The third experiment shows that, given a fixed number of average degree say five, namely on average each people has five friends to communicate. If people prefer communicating with local friends, then local communities are formed and so global consensus is hindered. In contrast, global consensus requires people to have sufficient chance to communicate globally.

## 4   Conclusions

In this paper, naming game (NG) is implemented by employing the multi-local-world (MLW) model, together with three typical topologies, namely random-graph, small-world and scale-free networks, as the underlying framework for communications. The underlying networks play an important role in NG, which indicate the relationships among different individuals, since connections are the precondition for pair-wise communications. As found in this study, community structures are essential for social communications, for which the MLW model used as the underlying network is more practical than the other commonly used network topologies.

The simulation is implemented to study the effects of the number and size of local-worlds in different NG networks, with or without communities, and the results are compared against several key parameters. Simulation results suggest that: 1) sufficiently many inter-community connections are crucial for the convergence; thus, given constant inter-connections, when intra-connections increase, meaning that the inter-connections are relatively weakened, the convergence process will be slowed down and eventually failed; 2) for sufficiently many inter-community connections, both the number and the size of communities do not affect the convergence at all; and 3) given the same average degree for different underlying network topologies, different clustering degrees will distinctively affect the convergence, which also change the shapes of the convergence curves.

The results of this investigation reveal the essential role of communities in NG on various complex networks, which shed new lights onto a better understanding of the human language development, social opinion forming and evolution, and even rumor epidemics alike.

**Acknowledgement**

This research was supported by the Hong Kong Research Grants Council under the GRF Grant CityU 11234916 and the National Natural Science Foundation of China under Grant No. 61473321.

**References**

[1]    Baronchelli, A., Felici, M., Loreto V., Caglioti, E., & Steels, L. Sharp transition towards shared vocabularies in multi-agent systems. *J Stat Mech: Theory Exp*, 6, P06014 (2006).

[2]    Lu, Q., Korniss, G., & Szymanski, B.K. The naming game in social networks: community formation and consensus engineering. *J Econ Interact Coord*, 4, 221–235 (2009).

[3]    Wang, W.X., Lin, B.Y., Tang, C.L. & Chen, G.R. Agreement dynamics of finite-memory language games on networks. *Eur Phys J B*, 60, 529–536 (2007).

[4]    Vylder, B.D., & Tuylsl, K. How to reach linguistic consensus: A proof of convergence for the naming game. *J Theor Biol*, 242, 818–831 (2005).

[5]    Centola, D., & Baronchelli, A. The spontaneous emergence of conventions: An experimental study of cultural evolution. *Proc Natl Acad Sci* (*PNAS*), 112, 1989–1994 (2015).

[6]    Erdős, P., & Rényi, A. On the strength of connectedness of a random graph. *Acta Math. Hungar*, 12, 261–267 (1961).

[7]    Watts, D.J., & Strogatz, S.H. Collective dynamics of 'small-world' networks. *Nature*, 393, 440–442 (1998).

[8]    Barabási, A., Albert, R., & Jeong H. Scale-free characteristics of random networks: the topology of the world-wide web. *Physica A*, 281, 69–77 (2000).

[9]    Baronchelli, A., Loreto, V., & Steels, L. In-depth analysis of the naming game dynamics: the homogeneous mixing case. *Int J Mod Phys C*, 19, 785–812 (2008).

[10]   Dall'Asta, L., Baronchelli, A., Barrat, A., & Loreto, V. Nonequilibrium dynamics of language games on complex networks. *Phys Rev E* 74, 036105 (2006).

[11]   Baronchelli, A., Dall'Asta, L., Barrat, A., & Loreto, V. The role of topology on the dynamics of the naming game. *Eur Phys J Special Topics*, 143, 233–235 (2007).

[12]   Dall'Asta, L., Baronchelli, A., Barrat, A., & Loreto, V. Agreement dynamics on small-world networks. *Europhys Lett*, 73, 969 (2006).

[13]   Liu, R.R., Jia, C.X., Yang, H.X., & Wang, B.H. Naming game on small-world networks with geographical effects. *Physica A: Statistical Mechanics and its Applications*, 388, 3615–3620 (2009).

[14]   Fu G., Zhang W. Naming game with biased assimilation over adaptive networks. *Physica A: Statistical Mechanics and its Applications*, 490, 260–268 (2018).

[15]   Fu G., Cai Y., & Zhang W. Analysis of naming game over networks in the presence of memory loss. *Physica A: Statistical Mechanics and its Applications*, 479, 350–361. (2017).

[16]   Lou Y., Chen G., & Hu J. Communicating with sentences: A multi-word naming game model. *Physica A: Statistical Mechanics and its Applications*, 490: 857–868 (2018)

[17]   Baronchelli, A. Role of feedback and broadcasting in the naming game. *Phys Rev E*, 83, 046103 (2011).

[18]   Li, B., Chen, G.R., & Chow, T.W. Naming game with multiple hearers. *Commun Nonlinear Sci Numer Simul*, 18, 1214–1228 (2013).

[19]   Gao, Y., Chen, G.R., & Chan, R.H.M. Naming game on networks: let everyone be both speaker and

hearer. *Sci Rep*, 4, 6149, doi:10.1038/srep06149 (2014).

[20]  Chen G.R., Fan, Z.P., & Li X. Modelling the complex internet topology. In *Complex Dynamics in Communication Networks*, Springer, 213-234 (2005).

[21]  Fan, Z.P., Chen G.R., & Zhang Y.N. A comprehensive multi-local-world model for complex networks. *Phys Letts A*, 373, 18, 1601-1605 (2009).

[22]  Siganos, G., Faloutsos, M., Faloutsos, P., & Faloutsos, C. Power laws and the AS-level internet topology. *IEEE ACM T Network*. 11, 514-524 (2003).

[23]  Lou Y. and Chen G.R. Analysis of the "naming game" with learning errors in communications. *Sci Rep*, 5, 12191; doi: 10.1038/srep12191 (2015).

[24]  Guo D.W., Meng X.Y., Liu M., & Hou C.F. Naming game on multi-community network (in Chinese with English Abstract). *J Comput Resear Devel* 52, 2, 487–498 (2015).

[25]  Lou Y., Chen G.R., Fan Z.P., & Xiang L.N. Supplementary Information for paper "Local communities obstruct global consensus: Naming game on multi-local-world networks": http://www.ee.cityu.edu.hk/~gchen/pdf/MLW-SI.pdf

# Supplementary Information for paper "Local communities obstruct global consensus: Naming game on multi-local-world networks"

Yang Lou[1], Guanrong Chen[1], Zhengping Fan[2*], and Luna Xiang[3]

[1] Department of Electronic Engineering, City University of Hong Kong, Hong Kong SAR, China
[2] School of Data and Computer Science, Sun Yat-sen University, Guangzhou 510275, China
[3] Department of Applied Social Sciences, City University of Hong Kong, Hong Kong SAR, China
*Corresponding author: fanzhp@mail.sysu.edu.cn

## Contents

## Abstract

This supplementary document further examines the scaling property of the population size (denoted by $N$) of 500 and 1500, respectively. *Convergence time* is used as the measure, which means the time steps needed to reach global convergence. The comparing simulation is carried out by varying parameters $\rho$, $m_0$ and $N_{LW}$. In the following comparison: 1) $\rho = 0.7$ is fixed and the *convergence time* affected by the dynamics of the number and size of the local-worlds is investigated; 2) the *convergence time* is studied when the rate $\rho$ of the initial assigned nodes is varied, while $m_0$ and $N_{LW}$ are all fixed; and 3) the convergence progresses of the multi-local-world (MLW) networks with the three typical models, namely random-graph (RG), small-world (SW) and scale-free (SF) networks, are compared.

# 1   Convergence time vs the number and size of local-worlds

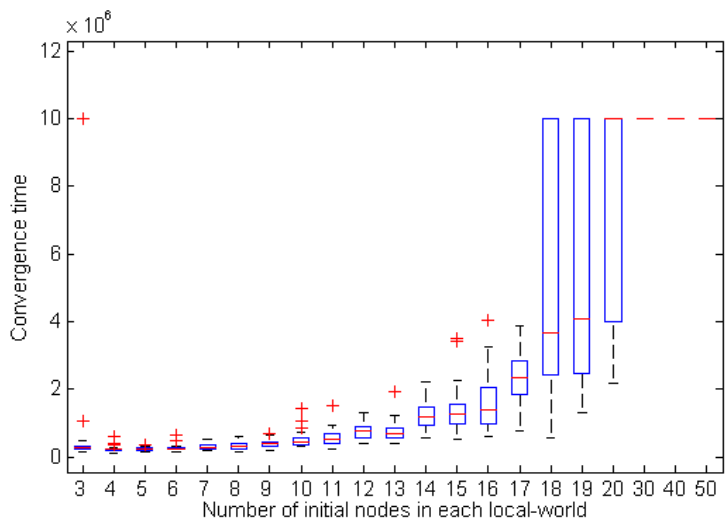

(a)   $N = 500$

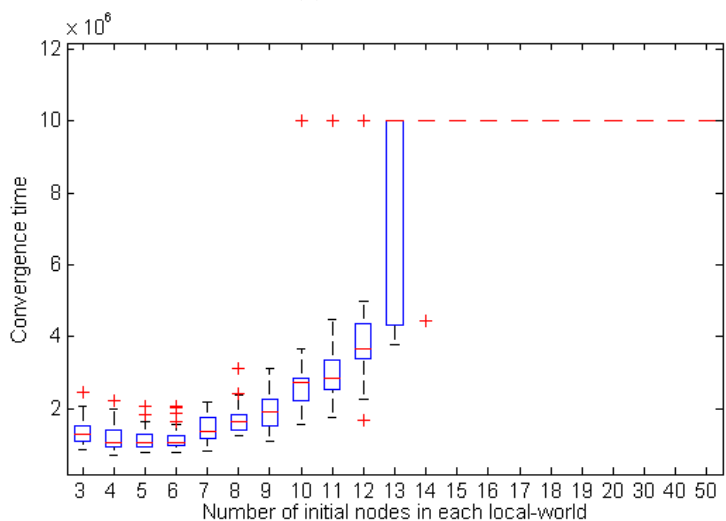

(b)   $N = 1500$

**Figure 1 The box plot of the convergence time vs the number $m_0$ of initial nodes in each local-world, with (a) $N = 500$ and (b) $N = 1500$.** In both cases, $\rho = 0.7$. The number of local-worlds is calculated by Eq. (4) in the paper, i.e., $N_{LW} = \lfloor \rho N/m_0 \rfloor$, where $\lfloor x \rfloor$ is to the largest integer less than or equal to $x$. In (a), $N = 500$, the mean value of the convergence time seemingly does not change as $m_0$ increases from 3 to 5, but it can be clearly observed that the maximum number of outliers (red plus) becomes smaller when $m_0$ increases from 3 to 5. In (b), $N = 1500$, both the mean value (red bar) and the maximum number of outliers (red plus) decrease when $m_0$ increases from 3 to 5, and then increases as $m_0$ continues to increase. Both box plots show a similar feature as that in the paper, where $N = 1000$: the overall convergence time firstly decreases as $m_0$ increases from 3 to 5, and then increases when $m_0 > 6$.

## 2 Convergence time vs the rate of initially assigned nodes

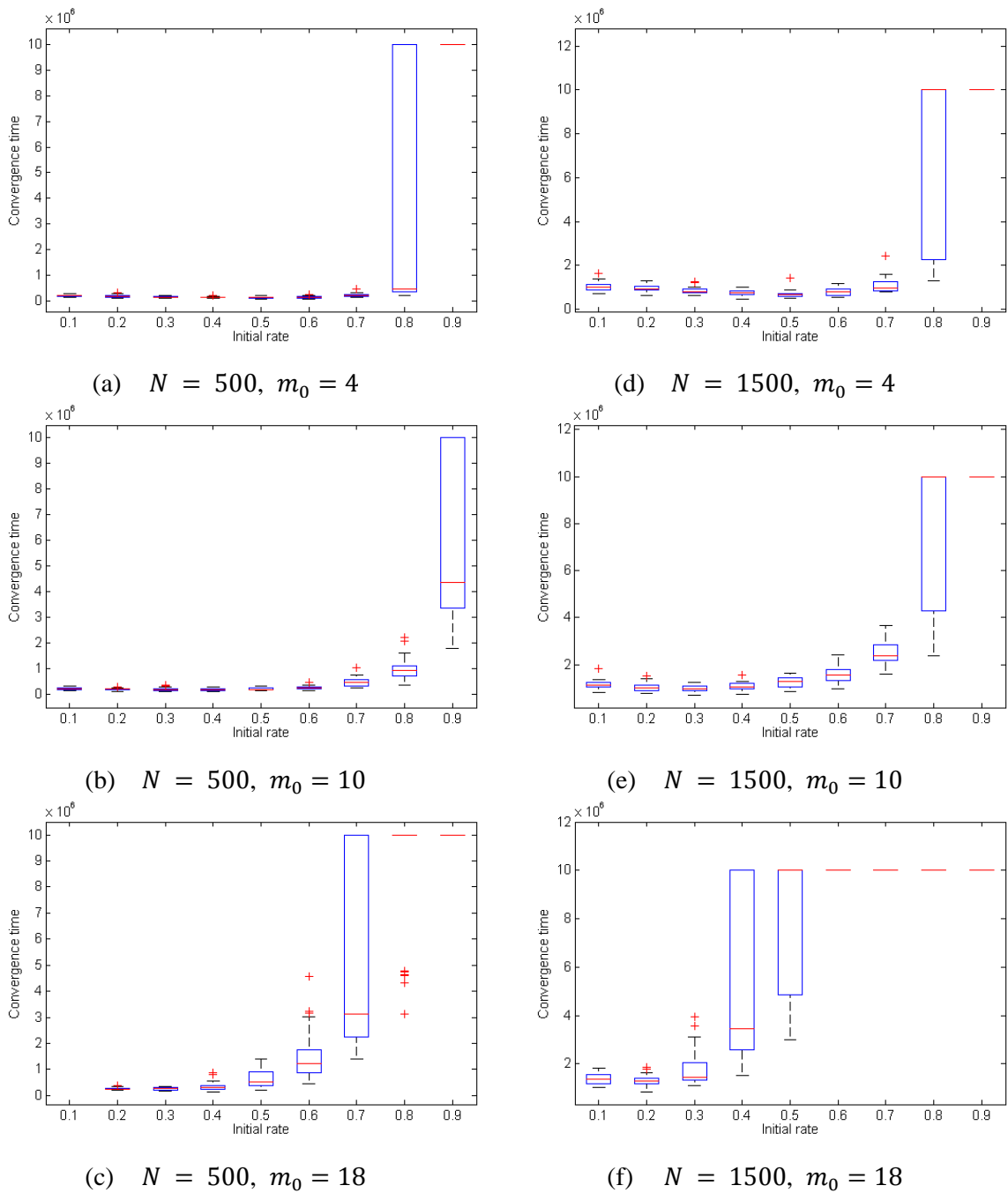

**Figure 2** **The box plot of the convergence time vs the rate $\rho$ of the initially assigned nodes.** The number $m_0$ of initial nodes in each local-world is set to 4, 10, and 18, respectively; the number of local-worlds is calculated by $N_{LW} = \lfloor \rho N / m_0 \rfloor$. In sub-figures (a), (b), (d), (e) and (f), $\rho$ is varied from 0.1 to 0.9, while in sub-figures (c), $\rho$ is varied from 0.2 to 0.9, to show different scenarios. The data for the case of $\rho = 0.1$ is missing because when $\rho = 0.1$, $N_{LW} = \lfloor \rho N / m_0 \rfloor = \lfloor 0.7 * 500 / 18 \rfloor = 2$, but both $N_{LW}$ and $m_0$ are supposed to be greater than or equal to 3. A common feature is that, when $\rho$ is small enough (i.e., $\rho \leq 0.7$ in sub-figures (a) and (d); $\rho \leq 0.6$ in (b); $\rho \leq 0.3$ in (c); $\rho \leq 0.5$ in (e); and $\rho \leq 0.2$ in (f)), the different values of $\rho$ do not affect the convergence time at all. However, when $\rho$ becomes greater than these values, the convergence time increases substantially.

## 3 Convergence processes

**Table 6 The feature statistics of the four networks, when $N = 500$.** Here, $\langle k \rangle$ is the average degree, $\langle pl \rangle$ is the average path length and $\langle cc \rangle$ is the average clustering coefficient. The data for MLW is collected from experiments in Section 1, while those for RG, SW and SF networks are generated using the $\langle k \rangle$ values of MLW for reference. As a result, the four types of networks have very similar $\langle k \rangle$ values.

| Reference | $\langle k \rangle$ | | | | $\langle pl \rangle$ | | | | $\langle cc \rangle$ | | | |
|---|---|---|---|---|---|---|---|---|---|---|---|---|
| | MLW | RG | SW | SF | MLW | RG | SW | SF | MLW | RG | SW | SF |
| $m_0 = 10$ | **9.40** | 9.40 | 10.00 | 9.92 | **3.52** | 3.01 | 3.43 | 2.76 | **0.63** | 0.02 | 0.34 | 0.08 |
| $m_0 = 20$ | **15.85** | 15.84 | 16.00 | 15.83 | **3.19** | 2.56 | 2.84 | 2.46 | **0.78** | 0.03 | 0.36 | 0.10 |
| $m_0 = 30$ | **22.30** | 22.33 | 22.00 | 21.72 | **3.00** | 2.31 | 2.59 | 2.28 | **0.82** | 0.04 | 0.38 | 0.12 |
| $m_0 = 50$ | **37.43** | 37.42 | 38.00 | 37.19 | **2.70** | 1.98 | 2.20 | 2.01 | **0.88** | 0.08 | 0.40 | 0.17 |

**Table 7 The feature statistics of the four networks when $N = 1500$.**

| Reference | $\langle k \rangle$ | | | | $\langle pl \rangle$ | | | | $\langle cc \rangle$ | | | |
|---|---|---|---|---|---|---|---|---|---|---|---|---|
| | MLW | RG | SW | SF | MLW | RG | SW | SF | MLW | RG | SW | SF |
| $m_0 = 10$ | **9.22** | 9.05 | 10.00 | 9.98 | **4.25** | 3.57 | 4.07 | 3.12 | **0.64** | 0.01 | 0.33 | 0.03 |
| $m_0 = 20$ | **16.10** | 15.75 | 16.00 | 15.94 | **3.69** | 2.92 | 3.40 | 2.77 | **0.78** | 0.01 | 0.37 | 0.05 |
| $m_0 = 30$ | **23.36** | 23.39 | 22.00 | 21.91 | **3.38** | 2.67 | 3.02 | 2.61 | **0.84** | 0.02 | 0.39 | 0.05 |
| $m_0 = 50$ | **37.35** | 37.23 | 38.00 | 37.73 | **3.22** | 2.36 | 2.64 | 2.30 | **0.90** | 0.02 | 0.38 | 0.08 |

In the following, Figures 3 and 6 show the convergence processes in terms of *the number of total words*, Figures 4 and 7 show the convergence processes in terms of *the number of different words*, and Figures 5 and 8 show the convergence processes of *the success rate*, with $N = 500$ and $N = 1500$, respectively. There are totally 24 sub-figures, and all these sub-figures show the same phenomenon that the blue plus (RG) converges the fastest, followed by the green triangles (SW), and the black stars (SF) ranks the third, and the red circles (MLW) always converge the slowest. This result supports the conclusions summarized in the paper.

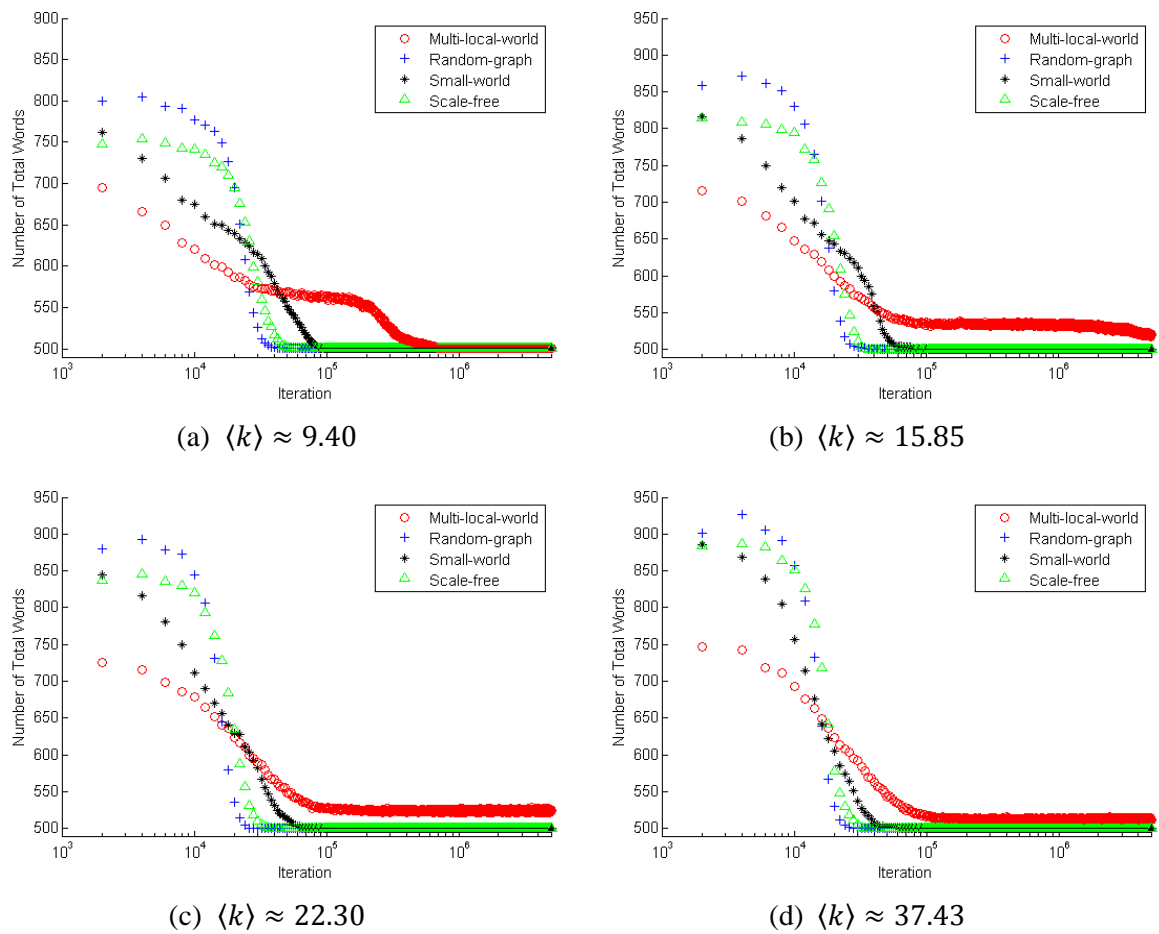

(a) $\langle k \rangle \approx 9.40$

(b) $\langle k \rangle \approx 15.85$

(c) $\langle k \rangle \approx 22.30$

(d) $\langle k \rangle \approx 37.43$

**Figure 3 Comparison of the convergence processes in terms of the number of total words ($N = 500$).**

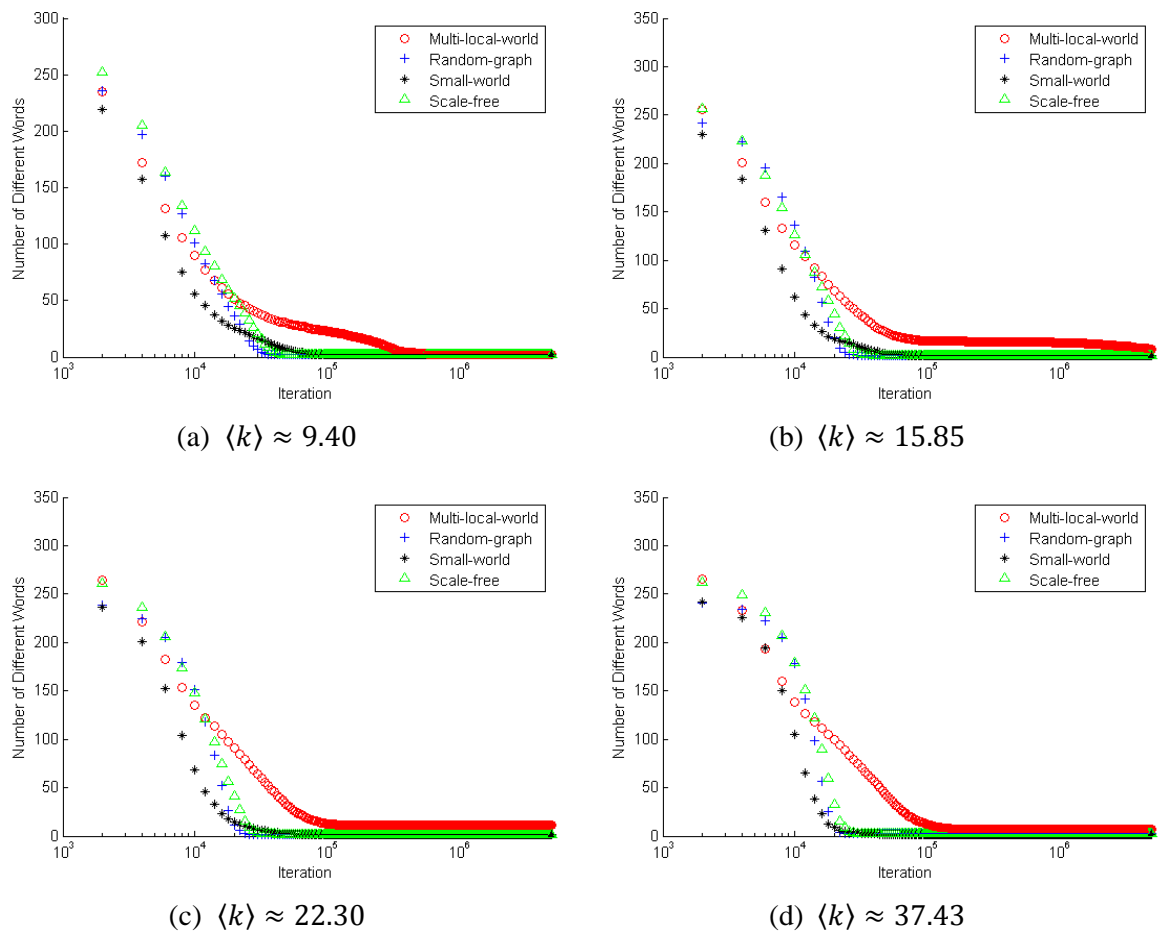

(a) $\langle k \rangle \approx 9.40$

(b) $\langle k \rangle \approx 15.85$

(c) $\langle k \rangle \approx 22.30$

(d) $\langle k \rangle \approx 37.43$

**Figure 4  Comparison of the convergence processes in terms of the number of different words ($N = 500$).**

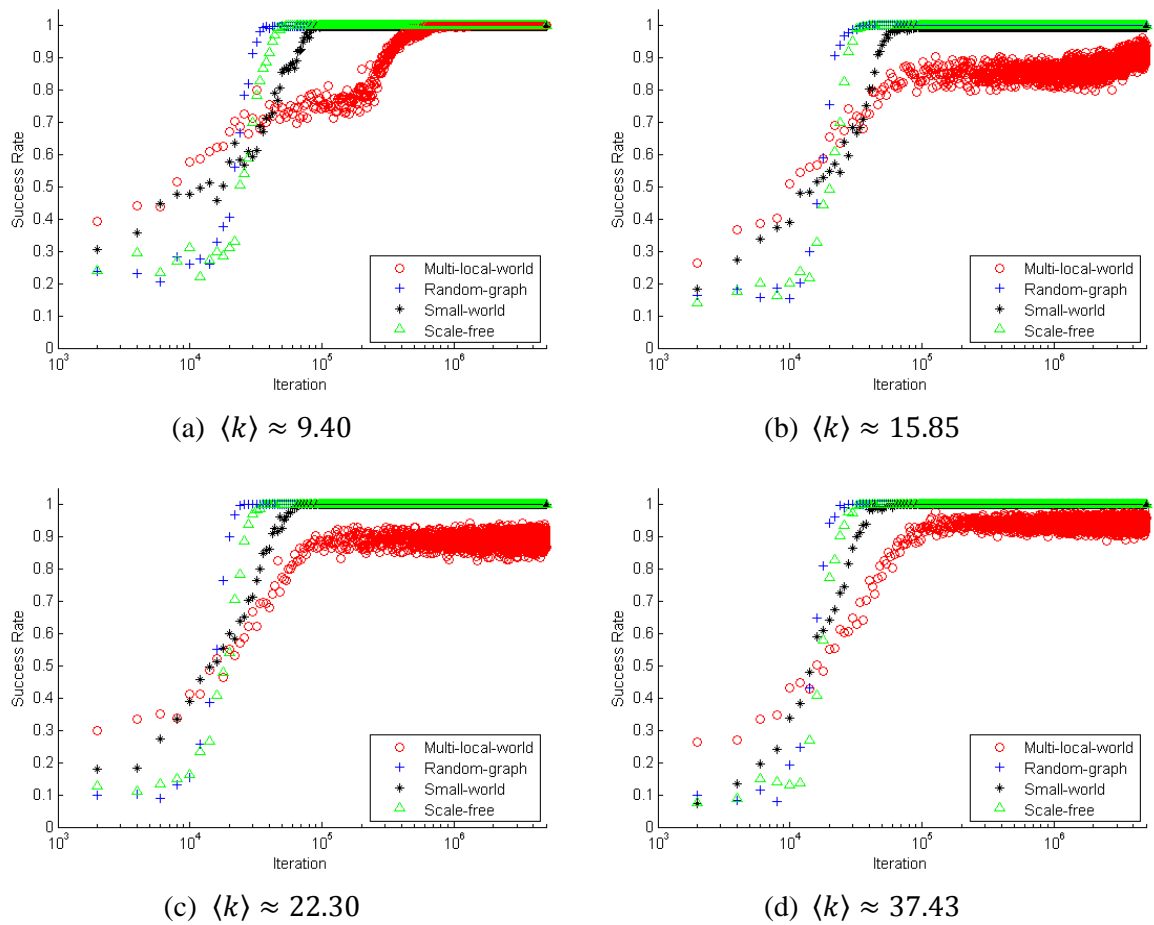

(a) $\langle k \rangle \approx 9.40$

(b) $\langle k \rangle \approx 15.85$

(c) $\langle k \rangle \approx 22.30$

(d) $\langle k \rangle \approx 37.43$

**Figure 5  Comparison of the convergence processes in terms of the success rate ($N = 500$).**

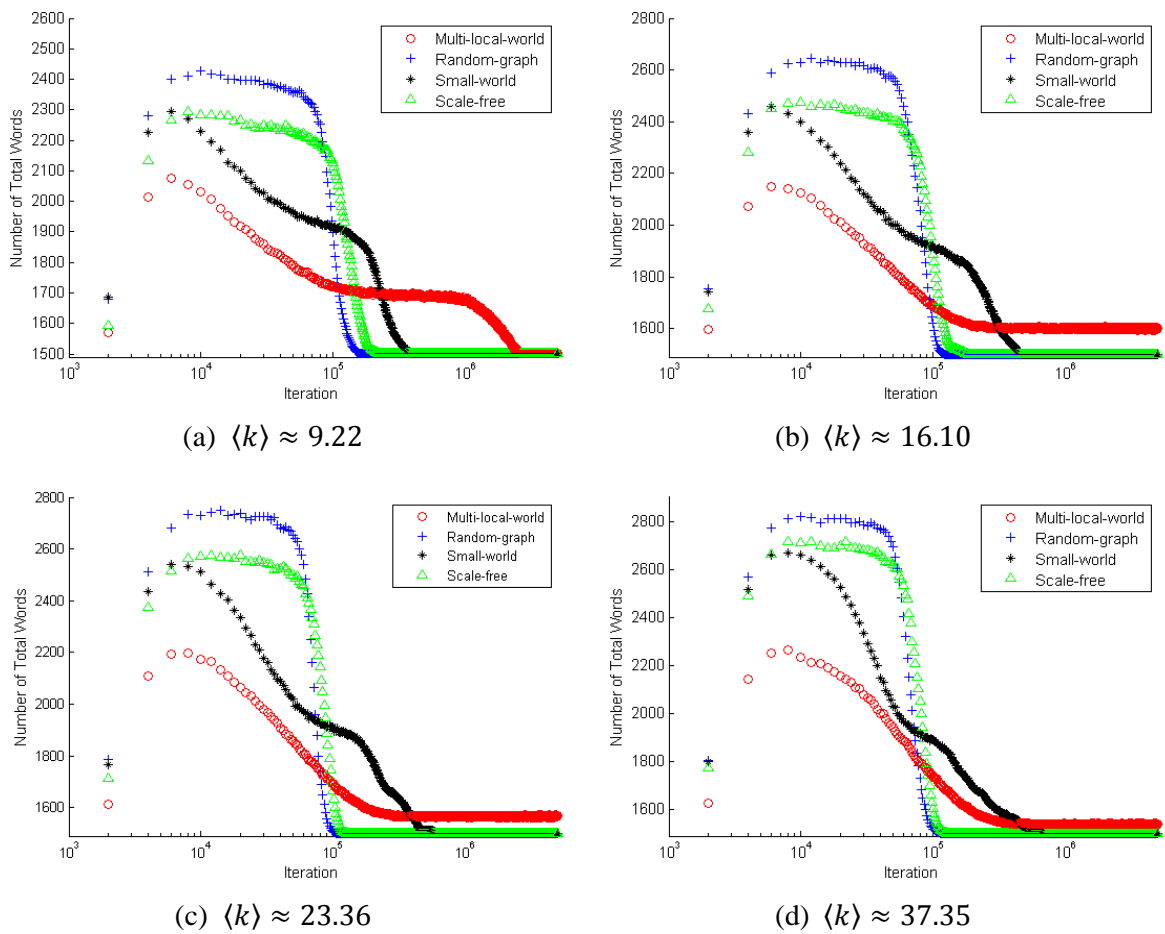

(a) $\langle k \rangle \approx 9.22$

(b) $\langle k \rangle \approx 16.10$

(c) $\langle k \rangle \approx 23.36$

(d) $\langle k \rangle \approx 37.35$

**Figure 6** **Comparison of the convergence processes in terms of the number of total words ($N = 1500$).**

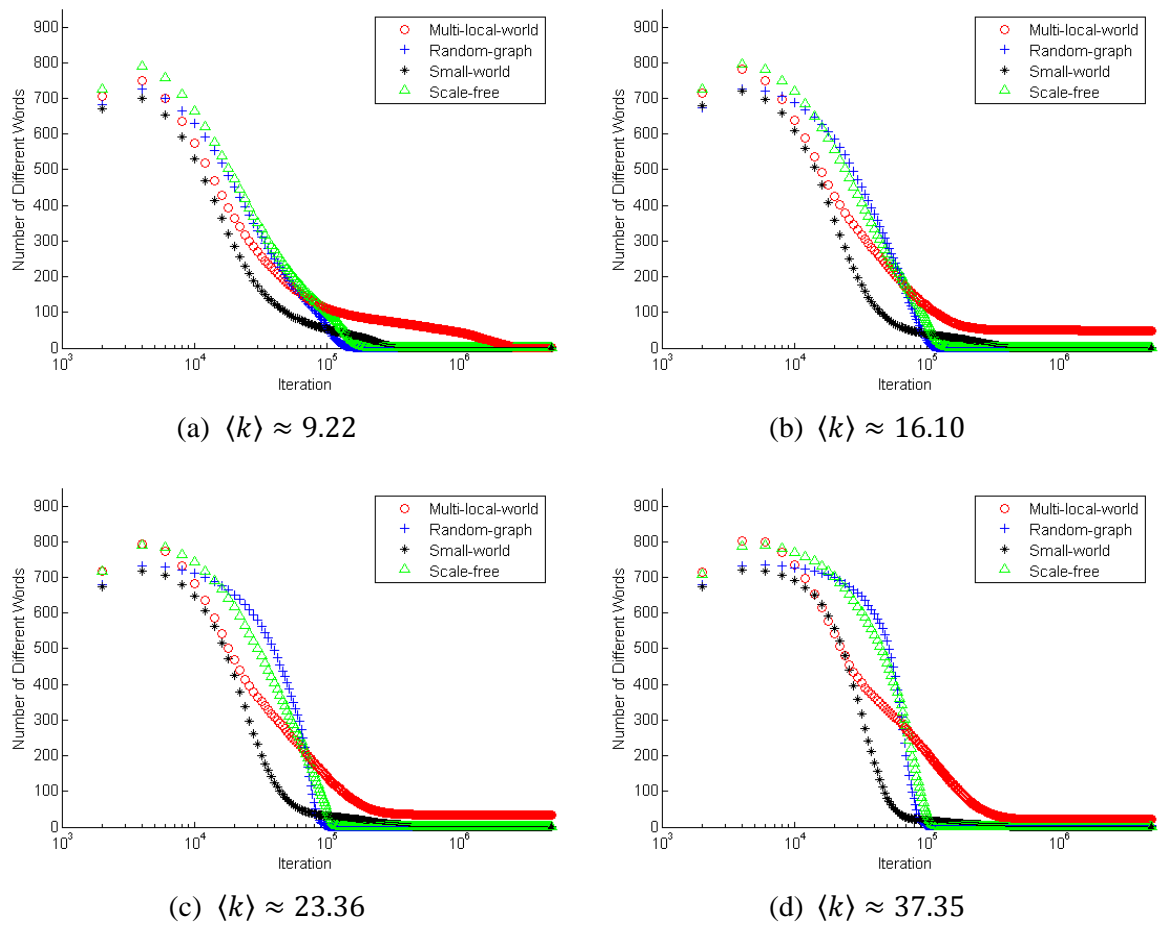

(a) $\langle k \rangle \approx 9.22$

(b) $\langle k \rangle \approx 16.10$

(c) $\langle k \rangle \approx 23.36$

(d) $\langle k \rangle \approx 37.35$

**Figure 7** **Comparison of the convergence processes in terms of the number of different words ($N = 1500$).**

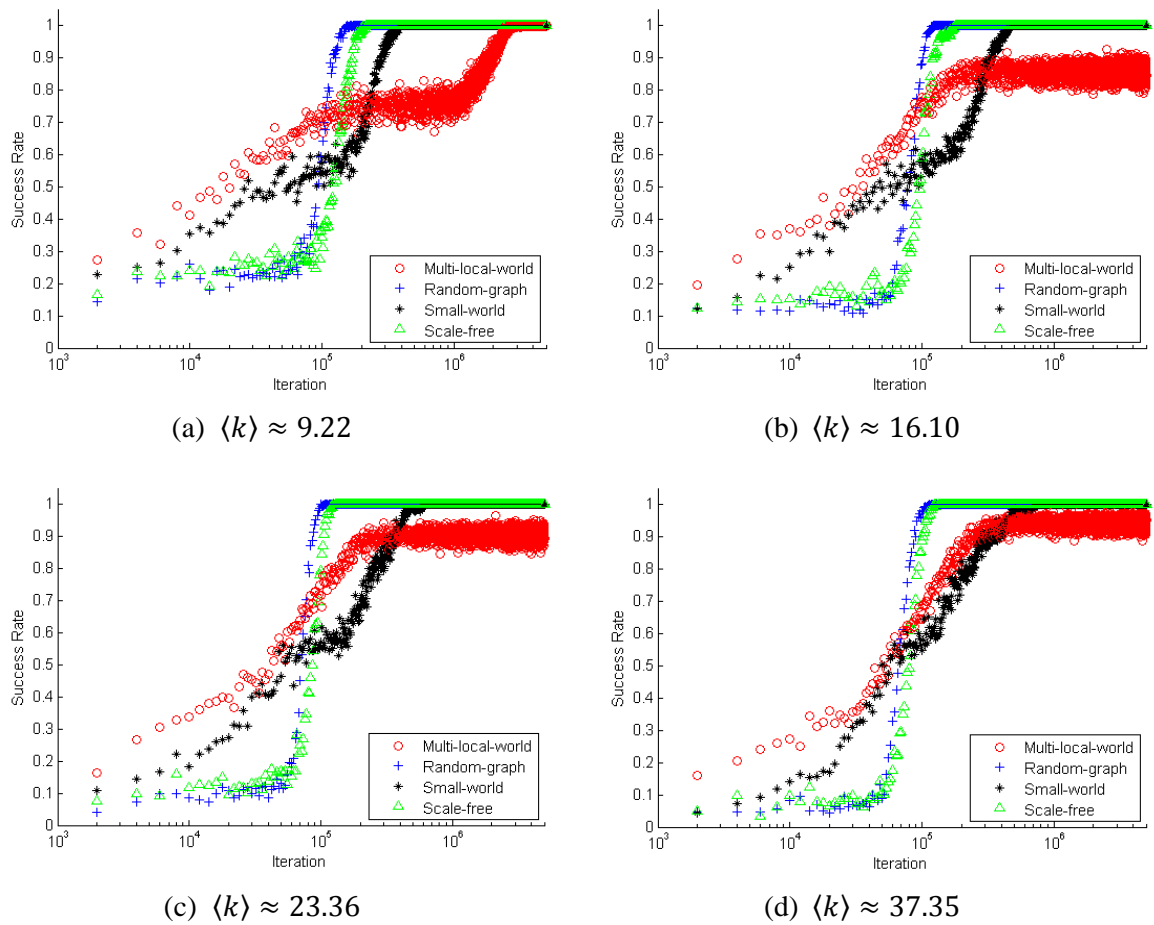

Figure 8  Comparison of the convergence processes in terms of the success rate ($N = 1500$).